\documentclass[twocolumn]{aastex701}
\usepackage{amsmath}
\usepackage{booktabs}

\begin{document}

\title{A TESS Search for Optical Variability in Intermediate-mass Black Hole Candidates}

\correspondingauthor{Jinghuai Yao, Wenwen Zuo,\\
and Hengxiao Guo}

\author[orcid=0009-0002-0525-8222]{Jinghuai Yao}
\affiliation{Department of Astronomy, University of Wisconsin-Madison,
Madison, WI 53706, USA}
\email[show]{jyao224@wisc.edu}

\author[orcid=0000-0002-4521-6281]{Wenwen Zuo}
\affiliation{Shanghai Astronomical Observatory, Chinese Academy of Sciences,
80 Nandan Road, Shanghai 200030, P. R. China}
\email[show]{wenwenzuo@shao.ac.cn}

\author[orcid=0000-0001-8416-7059]{Hengxiao Guo}
\affiliation{Shanghai Astronomical Observatory, Chinese Academy of Sciences,
80 Nandan Road, Shanghai 200030, P. R. China}
\email[show]{hengxiaoguo@gmail.com}

\author[orcid=0000-0002-3960-5870]{Ashutosh Tripathi}
\affiliation{State Key Laboratory of Radio Astronomy and Technology,
Xinjiang Astronomical Observatory, CAS, 150 Science 1-Street,
Urumqi 830011, China}
\email{ashutoshtripathi@xao.ac.cn}

\makeatletter
\renewcommand{\@@email}{%
jyao224@wisc.edu, wenwenzuo@shao.ac.cn,\\
hengxiaoguo@gmail.com}
\makeatother

\begin{abstract}

Optical variability of active galactic nuclei (AGNs) probes accretion onto black holes. We study the optical variability of AGNs in the mass range of intermediate-mass black holes (IMBHs, $10^{2-6}\,M_\odot$), whose expected variability damping timescales are of order days. We use the high-cadence photometry of the Transiting Exoplanet Survey Satellite (TESS) to search for optical variability within its individual 28-day sectors. By applying the \texttt{Quaver} light-curve reduction pipeline and designing artifact-aware variability metrics, we identify 3 sources with high-confidence variability and 17 sources with medium-confidence variability. Fitting damped random walk (DRW) models reveals a tentative timescale--mass correlation in a small, metric-selected subset, but not in the full sample. ZTF difference-image aperture photometry reveals long-term fractional variability of 1.8--7.4\% in the high-confidence sources, while most medium-confidence sources have variability below 1\%. This study demonstrates that the dense cadence and repeated full-sky observations of TESS are valuable for detecting IMBH optical variability, while also being limited by its instrumental artifacts and relatively shallow field.

\end{abstract}



\section{Introduction}

Intermediate-mass black holes (IMBHs) are typically defined as black holes (BHs) with masses between those of stellar-mass black holes (sBHs; $\lesssim 10^2\,M_\odot$) and supermassive black holes (SMBHs; $\gtrsim10^6\,M_\odot$). The demographic information of IMBHs is essential to our understanding of black-hole seeding, black-hole growth, and their occupation fraction in galaxies \citep{Greene20, Mezcua17}. 

The detection and confirmation of IMBHs has been rather elusive compared with sBHs and SMBHs: SMBHs can be detected through accretion signatures in active galaxies and through stellar or gas dynamics even in quiescent galaxies, whereas nearby sBHs can show up in X-ray binaries or compact-object mergers; IMBHs detection is challenging because their gravitational spheres of influence are smaller and their accretion signatures can be weak relative to the host galaxy. Over the past two decades, optical spectroscopy has established the first substantial samples of IMBH candidates. Broad-line searches in Sloan Digital Sky Survey data uncovered large samples of low-mass type~1 AGNs with virial black-hole masses below a few $10^{6}\,M_{\odot}$ \citep{Greene07}. Targeted studies of dwarf and low-mass galaxies identified additional candidates through narrow-line diagnostics, high-ionization lines, and persistent broad Balmer emission \citep{Reines13,Chilingarian18}. More recently, Dark Energy Spectroscopic Instrument (DESI) has also provided a large catalog of dwarf AGN candidates \citep{Pucha25, Pucha26}.

Optical spectroscopy often yields IMBH candidates instead of confirmed objects, usually because of contamination from stellar processes \citep{Baldassare16} and uncertainty in virial mass estimates \citep{Shen13, DallaBonta25}. In addition, optical spectroscopy may miss low-luminosity AGNs accompanied by star-formation features \citep{Baldassare20, Ward22, Yuk22}, which is indeed common in dwarf galaxies. As an alternative pathway, variability selection directly monitors the stochastic change of AGN emission. Variability-selection studies have used optical/near-infrared surveys including the PTF \citep{Baldassare20}, ZTF \citep{Ward22}, ASAS-SN \citep{Yuk22}, Kepler \citep{Smith18}, and TESS \citep{Burke20, Treiber23, Smith23}. 

Although designed for exoplanet discovery, TESS delivers nearly continuous light curves at cadences of approximately 3--30 minutes over 28-day sectors for most of the sky \citep{Ricker15}, in contrast to cadences of a few days of ground-based surveys. The high cadence makes TESS a powerful probe of variability on timescales from hours to days. AGN optical variability is often characterized by the damping timescale \(\tau_{\rm DRW}\) of a damped random walk model (DRW). Empirical studies find that this timescale increases with black-hole mass, with IMBHs expected to vary on day-to-week timescales that are well matched to TESS sector baselines \citep{Burke21,Wang23}.

On the other hand, TESS has a broad point-response function (PRF), a relatively shallow field, and residual instrumental artifacts that can mimic weak AGN-like variability \citep{Vanderspek18}. \cite{Smith23} introduced \texttt{Quaver} to address these instrumental issues and extract TESS light curves specifically for AGNs. Aperture selection and background correction in \texttt{Quaver} require extensive manual inspection for each source in each sector. In this study, we automate \texttt{Quaver} light curve retrieval and design variability-detection metrics followed by manual inspection; this procedure makes \texttt{Quaver} compatible with large, systematic surveys. While previous studies focus on the primary mission of TESS from 2018 to 2020, we include TESS observations spanning a period of 8 years, with sectors of different cadences normalized to the same standard. Most of the IMBH candidates are observed in 3 sectors or more, enabling us to evaluate the change of variability over time and increase the confidence of variability selection.

The paper is organized as follows. In Section~\ref{sec:data}, we describe the parent IMBH candidate catalog and the TESS observations. Section~\ref{sec:methods} presents the \texttt{Quaver} light-curve reduction, our variability-detection metrics, and the variability timescale modeling techniques. Section~\ref{sec:results} summarizes the survey-level results, presents the timescale analysis, and examines long-term variability. In Section~\ref{sec:conclusion} we discuss the implications for IMBH searches and for the use of TESS as a variability survey for low-mass accretors.

\section{Data and Sample Construction}\label{sec:data}

\subsection{IMBH candidate catalog}

For the candidate catalog, we combined the two recent low-mass AGN candidate catalogs. The first one is from \cite{Guo26}, which assembles IMBH candidates and derives homogeneous virial mass estimates from broad H$\alpha$ emission; the second one is the dwarf AGN catalog derived from DESI DR1 \cite{Pucha25, Pucha26}. For sources common to both catalogs, we adopt the black hole mass estimate from \citet{Guo26} in the result section. If the reported statistical mass uncertainty is $\geq1$ dex, we instead adopt the archival mass compiled by \citet{Guo26}.

The target selection is based on black hole mass and source brightness. In this work, we adopt an IMBH-candidate threshold of
$M_{\rm BH}<10^{6}\,M_{\odot}$ and retain a target when at least one available mass estimate satisfies this threshold, even if other estimates exceed it. To match the bandpass and field depth of TESS, we select sources using Zwicky Transient Facility (ZTF, \citealt{Bellm19}) photometric magnitudes, requiring $i_{\rm ZTF}<17.5$ as the primary cut. For sources without available $i$-band measurements, we use $r_{\rm ZTF}<17.8$ and $g_{\rm ZTF}<18.5$ as the first and second fallbacks, respectively. These band-dependent thresholds are determined by the average ZTF color offsets in the parent catalog, where r-band and g-band are 0.3 and 1.0 magnitudes fainter on average relative to i-band. A total of 410 sources meet the criteria for this study.

We also explored the variability of dwarf galaxy nuclei that are not cataloged as AGNs. We applied the same ZTF source brightness threshold to the DESI Extragalactic Dwarf Galaxy Catalog \citep{DESI_ExtragalacticDwarfs_v1, DESI2026_DR1}. To complement sky coverage beyond DESI, we also filtered nearby galaxies with stellar mass $M_{\rm stellar}\leq10^{9.5}\,M_{\odot}$ from the 50 Mpc Galaxy Catalog \citep{Ohlson24}. The stellar mass threshold is motivated by the mean stellar-to-black-hole mass ratio \citep{Reines15}, and a brighter threshold is chosen at $i_{\rm ZTF}<15.5$ because of the substantially larger source number. A combined total of 155 sources are examined using \texttt{Quaver}. Among these sources not cataloged as AGNs, we did not independently identify any high-confidence AGN variability candidates, and their results are not included in Section \ref{sec:results}.

\subsection{TESS observations}

TESS observes in a broad red-optical to near-infrared bandpass spanning approximately 600--1000\,nm. It has four cameras observing simultaneously, each with a $24^\circ \times 24^\circ$ field of view; together, they provide a $24^\circ \times 96^\circ$ field of view for each sector \citep{Ricker15}. In this work, we use the TESS full-frame image (FFI) products rather than pre-selected target-pixel light curves, allowing us to extract light curves for all IMBH candidates with available sector coverage. Each sector spans approximately 28 days, but the FFI cadence has changed significantly over the mission. The primary TESS mission from 2018 to 2020 (Sectors 1--26) obtained FFIs at 30 min cadence; the first extended mission from 2020 to 2022 (Sectors 27--55) used 10 min cadence; and the second extended mission (Sector 56 onward) uses 200 s cadence. The calibrated FFIs are available from MAST at
\dataset[doi:10.17909/0cp4-2j79]{https://doi.org/10.17909/0cp4-2j79}.

\begin{figure}[htbp]
\centering
\includegraphics[width=0.45\textwidth]{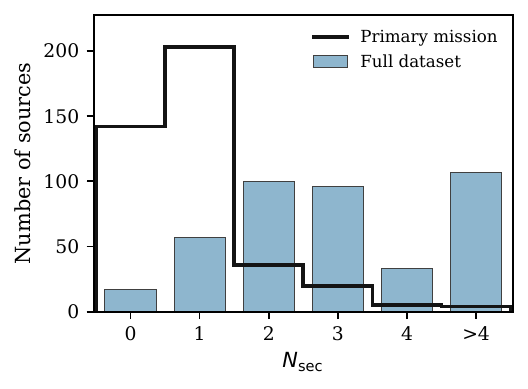}
\caption{Distribution of the number of available TESS sectors per IMBH candidate. The filled histogram shows the coverage of the full dataset used in this study, while the black outline shows the coverage from the primary mission only.
\label{fig:Nsec}}
\end{figure}

While most TESS AGN variability studies have relied on primary-mission data \citep{Burke20,Treiber23,Yuk26}, we incorporate FFIs from the extended missions and perform median smoothing to account for the cadence differences. As shown in Figure \ref{fig:Nsec}, incorporating the extended missions improves the coverage of observed candidates (primary: 65\%; full: 96\%) and significantly increases the chance of multi-sector revisits (primary: 16\%; full: 82\%). This allows us to analyze TESS sectors spanning an 8-year period (Cycles 1--8; Sectors 1--107), enabling variability detection at multiple epochs, which is particularly useful for sources with intermittent variability. In Section~\ref{sec:results}, we demonstrate that the detection statistics do not show a significant trend across different mission phases after proper normalization.

\subsection{ZTF observations}

Because TESS light curves measure relative flux and often have long gaps between observations, we use publicly available Zwicky Transient Facility data (ZTF; \citealt{Bellm19,Masci19}) from DR24, spanning March 2018 to October 2025, for source brightness screening and long-term variability assessment; it is also used to monitor neighboring sources that may mimic AGN variability in TESS light curves. The ZTF images and light curves are available from IRSA at \dataset[doi:10.26131/IRSA539]{https://doi.org/10.26131/IRSA539}
and \dataset[doi:10.26131/IRSA598]{https://doi.org/10.26131/IRSA598}, respectively.

For source brightness screening, we estimate the source magnitude from the median of the 9 most recent PSF-photometry measurements and apply the band-dependent brightness cuts described in Section~2.1. 

For long-term variability analysis, we performed forced aperture photometry on ZTF difference images \citep{Arevalo26, Zuo26}. We use the $r$ band, which provides the densest temporal coverage for our targets. The forced photometry is performed at an aperture of 3 pixels radius using \texttt{SExtractor} \citep{Bertin96} in dual-image mode, with the science image used for source detection and the difference image for flux measurement. Because some sources have limited epochs available in the same ZTF fields, we combined data from different fields by assuming a constant offset between fields. We require at least 20 overlapping nights to estimate the offset between fields; fields with fewer overlapping nights are excluded. After combining same-night measurements, we applied two iterations of $5\sigma$ clipping and then excluded any night differing from the median of other nights within $\pm20$ days by more than $3\sigma$.

\section{Methodology}\label{sec:methods}

\subsection{TESS light curve reduction: \texttt{Quaver}}

We reduce the TESS FFIs with a survey-mode adaptation of \texttt{Quaver}. \texttt{Quaver} is an AGN-oriented pipeline designed to extract and detrend TESS light curves for stochastic nuclear variability \citep{Smith23}.\footnote{\url{https://github.com/kristalynns/Quaver}} \texttt{Quaver} models the dominant instrumental trends using regressors built from a $35 \times 35$ pixel cutout from each FFI. In this work we use the full-hybrid mode correction, in which the aperture light curve is regressed against additive principal components from faint background pixels, second-order additive terms, and multiplicative principal components derived from bright pixels after additive correction. Note that TESS light curves reduced by \texttt{Quaver} are not suitable for absolute photometry \citep{Smith23}.

\begin{figure*}[htbp]
\centering
\includegraphics[width=0.8\textwidth]{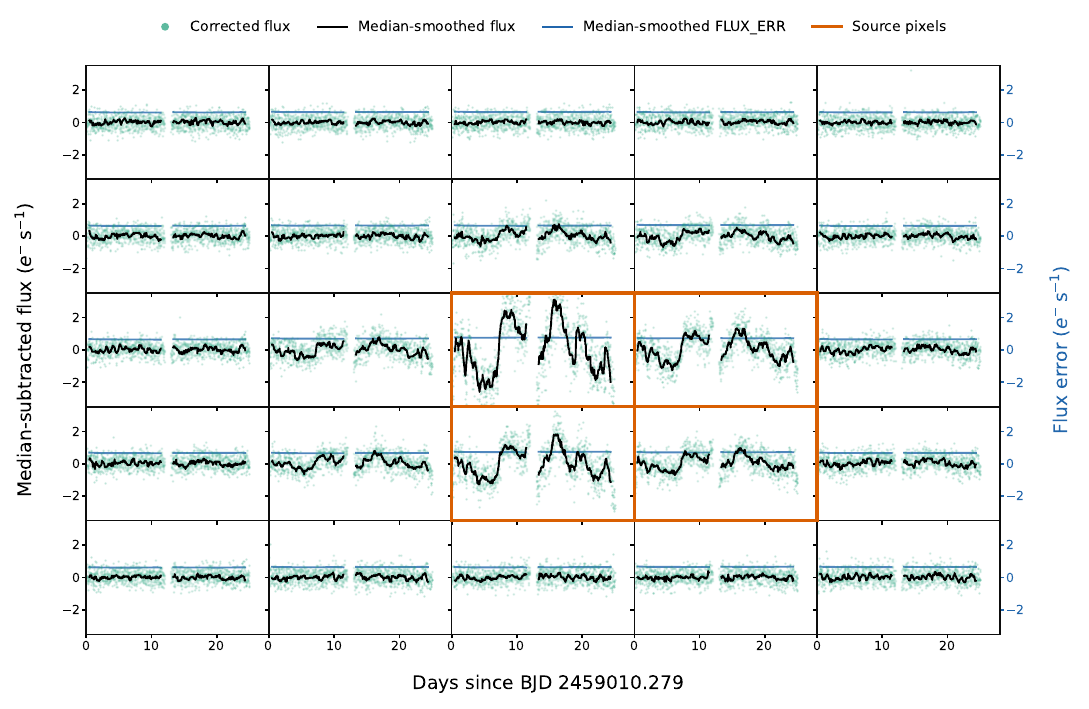}
\caption{Example \(5\times5\) pixel-grid product for J114515.94+794053.4 Sector 26 after \texttt{Quaver} reduction. Green points show corrected fluxes, whereas black and blue curves show the 0.5-day median-smoothed flux and flux error. The four source pixels are outlined in orange.
\label{fig:Flux_grid}}
\end{figure*}

We modified \texttt{Quaver} to operate in survey mode to process IMBH candidates in large numbers. The automated approach accepts target lists or FITS catalog tables, automatically masks the flux spikes caused by instrumental artifacts, and extracts light curves in a $5 \times 5$ grid centered at the source coordinate. Following the visual inspection approach of \citet{Treiber23}, the grid products shown in Figure~\ref{fig:Flux_grid} enable direct comparison of the central-source light curve with those of neighboring
background pixels, providing the basis for the variability-confidence metrics described below.

\subsection{Variability detection metrics}\label{ssec:metrics}

For variability detection, we do not compute the S/N of light-curve variations directly from the nominal flux error reported by TESS. Although \texttt{Quaver} removes much of the instrumental variability present in TESS FFIs, residual artifacts and contamination from neighboring sources can still mimic AGN variability, and these features are not well captured by the flux error. Therefore, we design custom variability detection metrics to filter the candidates with potential intrinsic variability, and then manually inspect the targets to evaluate the confidence of detection.

To distinguish intrinsic variability from residual artifacts, we design two metrics to quantify the variability detection confidence: amplitude confidence ($C_{\mathrm{AMP}}$) and correlation confidence ($C_{\mathrm{CORR}}$). Motivated by the visual inspection approach of \citet{Treiber23}, these metrics quantify the source-to-background amplitude contrast and relative light-curve coherence, enabling systematic screening of large samples.

\subsubsection{Amplitude confidence}

The amplitude confidence ($C_{\mathrm{AMP}}$) evaluates how strongly the source varies compared to the background. The source pixels are defined as the four TESS pixels that receive the most source flux assuming a PRF model (see Figure \ref{fig:Flux_grid}).

The light curves of source pixels are smoothed with a window of 0.5 days\footnote{The smoothing suppresses cadence-scale noise across different missions while retaining day-scale variability. Tests with 0.25-day and 1.0-day windows give qualitatively consistent rankings.} and summed to form a compact source-aperture light curve. The median smoothing address the different noise levels among sectors observed at different cadences, effectively normalizing data acquired in different TESS missions.
The variability amplitude ($A_{src}$) is measured using the biweight scale, a robust scatter estimator that downweights outlying cadences.

The global background amplitude ($A_{\mathrm{bg,global}}$) is defined as the median amplitude of four-pixel apertures in the background of the $5 \times 5$ grid. Because residual instrumental artifacts can be enhanced along specific detector columns, we also compute a source-column background amplitude ($A_{\mathrm{bg,column}}$) using only background pixels in the same columns as the source aperture. We adopt the larger of these two quantities as the resulting background amplitude. The amplitude confidence ($C_{\mathrm{AMP}}$) is then defined as the ratio of source amplitude to background amplitude:
\[
C_{\mathrm{AMP}} =
\frac{A_{\mathrm{src}}}
{\max(A_{\mathrm{bg,global}}, A_{\mathrm{bg,column}})} .
\]
For J114515.94+794053.4 Sector 26 shown in Figure \ref{fig:Flux_grid}, the source-pixel light curves are significantly more variable than the background ($C_{\mathrm{AMP}}=9.88$), and the amplitude is also much higher compared to the per-cadence flux error.

\subsubsection{Correlation confidence}

The correlation confidence ($C_{\mathrm{CORR}}$) evaluates whether the source pixel light curves vary coherently with one another, rather than tracing variability that is also present in the background pixels. Ideally, real AGN variability should exhibit high correlation between source pixels, and near-zero correlation between source and background.

The source correlation ($r_{\mathrm{src}}$) is defined as the median Pearson correlation coefficient between each source pixel and the sum of the other three source pixels. The background correlation ($r_{\mathrm{bg}}$) is the mean absolute value of correlation between each background pixel and sum of all four source pixels, since some instrumental systematics can cause anti-correlated light curves. The correlation confidence is then defined as
\[
C_{\mathrm{CORR}} =
r_{\mathrm{src}} - r_{\mathrm{bg}}
\]
For J114515.94+794053.4 Sector 26 shown in Figure \ref{fig:Flux_grid}, the four source pixels are highly correlated, and the obtained $C_{\mathrm{CORR}}$ is 0.71. Combined with the high $C_{\mathrm{AMP}}$, this suggests that the significant variability comes from intrinsic AGN activity at the source position.

\subsubsection{Artifact and contamination screening}\label{sssec:aux}

The amplitude and correlation confidences serve as the primary metrics to identify likely variable sources, but they may fail when the light curves are contaminated by neighboring sources or severe instrumental artifacts. We therefore compute quality flags to address these circumstances. The TESS artifact quality flag is detailed in Appendix \ref{App:flag}. In the following sections, flagged sectors are excluded when selecting the representative largest-amplitude sector for each source and are not used for variability timescale modeling.

We also visually inspect contamination from neighboring sources, as the broad PRF may allow the variability of neighboring variable stars or quasars to leak into central source pixels. 
Therefore, we exclude targets with significantly brighter neighboring sources within the \(5\times5\) grid and note cases where nearby neighbors have comparable brightness.

\subsubsection{Variability timescale modeling}

TESS offers light curve monitoring at a unique hours-to-month timescale, matching the optical variability timescale of IMBHs. Therefore, we further model the highest-quality TESS sectors with a damped random walk (DRW), a commonly used model for stochastic optical AGN variability \citep[e.g.,][]{Kelly09,MacLeod10,Burke20}.

For variability timescale modeling, we sum the source-aperture \texttt{Quaver} flux without median smoothing and assume the per-cadence uncertainty to be the TESS flux errors added in quadrature.
We fit the DRW model using \texttt{EzTao} \citep{EzTao}, which implements the DRW as a CARMA(1,0) Gaussian process and models $\tau_{\rm DRW}$, the timescale over which the light curve loses memory of its previous state. In frequency space, the corresponding DRW PSD has the form
\[
P_{\rm DRW}(f) \propto
\frac{1}{1+\left(2\pi f \tau_{\rm DRW}\right)^2},
\]
and the DRW frequency is \(f_{\rm DRW}=1/(2\pi\tau_{\rm DRW})\), marking the transition between the low-frequency plateau and the high-frequency \(f^{-2}\) regime. We note that $\tau_{\rm DRW}$ values should be treated cautiously when they approach the sector duration \citep{Kozlowski17}.

The broken-power-law (BPL) PSD model is also commonly used to characterize AGN optical PSDs and their break timescales \citep[e.g.,][]{Smith18, Yuk22, Yuk25}. For BPL modeling, we compute Lomb--Scargle periodograms of the TESS light curves \citep{Lomb76,Scargle82} and fit the binned periodogram with \(P_{\rm BPL}(f)=P_{\rm floor}+A/[1+(f/f_{\rm br})^\alpha]\), where \(f_{\rm br}\) is the break frequency. However, the BPL model has more free parameters than the DRW model, making it more vulnerable to noise and residual artifacts, and robust constraints often require broader frequency coverage from multi-instrument or multi-baseline data. We therefore use the BPL only as a consistency check on the DRW result and do not use its parameters in source-property correlations. 

We quantify long-term variability by the fractional rms amplitude, $F_{\rm var}=\sqrt{S^2-\langle\sigma_F^2\rangle}/\langle F\rangle$,
where $S^2$, $\langle\sigma_F^2\rangle$, and $\langle F\rangle$ are the variance, mean squared uncertainty, and mean of the nightly fluxes, respectively. This measures the noise-corrected RMS variability relative to the mean flux \citep{Vaughan03, Simm15}.

\section{Results}\label{sec:results}

\begin{figure}[htbp]
\centering
\includegraphics[width=0.45\textwidth]{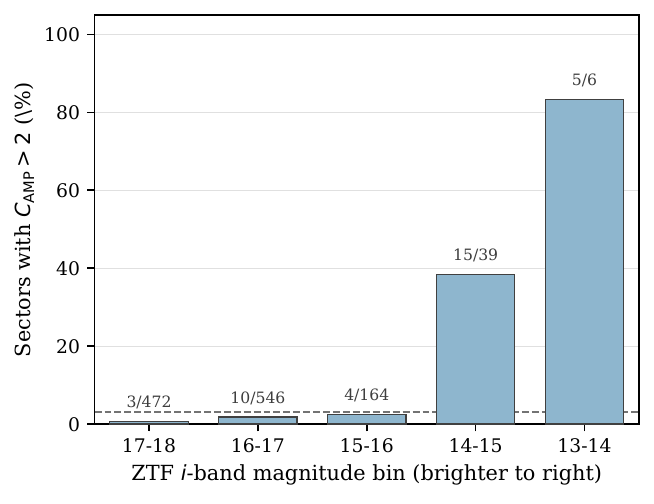}
\caption{Fraction of sectors with $C_{\mathrm{AMP}}>2$ versus ZTF $i$-band magnitude. Labels give the number above the threshold relative to all sectors in each bin; the dashed line marks the 3.0\% fraction among sectors with measured $i$-band magnitudes.
\label{fig:frac}}
\end{figure}

\subsection{Survey overview and variability detections}

Among the full sample, only 44 of 1457 sectors (3.0\%) have $C_{\mathrm{AMP}}>2$, and the brighter sources tend to have a larger fraction of $C_{\mathrm{AMP}}>2$, as shown in Figure \ref{fig:frac}. This implies that the variability detection of many faint sources is magnitude-limited by TESS. This is partly because the optical continua of low-mass/IMBH AGNs are often dominated by host-galaxy starlight, which dilutes the observed amplitude of nuclear variability in unresolved photometry \citep{Dong12, Arevalo26, Zuo26}. No significant trend can be confirmed with black hole mass, which is consistent with the lack of an amplitude--mass correlation reported by \citet{Smith18}. High amplitudes are detected down to $\log(M_{\rm BH}/M_\odot)=4.5$, and the black holes with even lower masses constitute only a small fraction of the catalog sample. Therefore, variability detection of lower mass IMBHs remains possible in future surveys. The metric distribution does not change noticeably across missions with different FFI cadences, which suggests that the median smoothing approach is effective in normalizing TESS data in different missions.

\begin{figure}[htbp]
\centering
\includegraphics[width=0.45\textwidth]{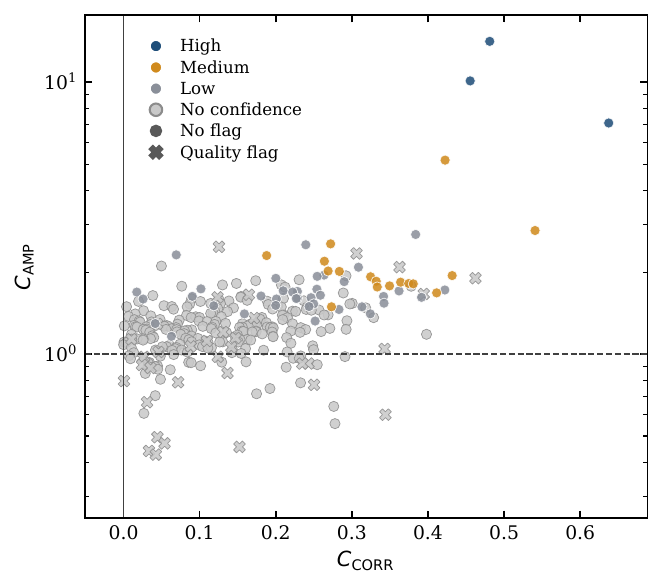}
\caption{Distribution of the amplitude confidence $C_{\rm AMP}$ and the correlation confidence $C_{\rm CORR}$, with the manual confidence classification labeled. Each source is represented by its largest $C_{\rm AMP}$ sector with $C_{\rm CORR}>0$. Crosses mark sources for which all available sectors are flagged.
\label{fig:manual}}
\end{figure}

Based on the variability metrics and the diagnostics defined in Section \ref{ssec:metrics}, we manually inspected the sources and determined the subjective confidence of IMBHs having optical variability. The confidence levels are qualitatively ranked high, medium, and low. Table~\ref{tab:high_medium_candidates} reports the metrics for a representative sector of each high- and medium-confidence target, chosen as the sector with the largest $C_{\rm AMP}$ among those with $C_{\rm CORR}>0$. Figure \ref{fig:manual} shows the distribution of sources with different manual confidence on the variability metric plane. A total of 3 high-confidence sources and 17 medium-confidence sources are identified. All three high-confidence sources, J114515.94+794053.4 (UGC 6728), J120309.60+443152.7 (NGC 4051), and J122548.87+333248.7 (NGC 4395), have independent literature evidence for optical variability (see \cite{Bentz16, Fausnaugh17, Burke20}, respectively). Among the medium-confidence candidates, only J172005.27+535703.0 is found in a photometric variability-selected AGN catalog \citep{Baldassare20}. 
The remaining medium-confidence candidates do not appear to have published optical-variability confirmations, and are therefore promising targets for dedicated follow-up.

\begin{table*}[htbp]
\small
\caption{Variability metrics and source properties for high- and
medium-confidence variable candidates.}
\label{tab:high_medium_candidates}

\par
\begingroup
\centering
\movetableright=0pt
\begin{tabular}{llcccccccc}
\toprule
Source & Class & $z$ & $\log(M_{\rm BH}/M_\odot)$ &
Sector & $N_{\rm sec}$ & $C_{\rm AMP}$ & $C_{\rm CORR}$ &
$i_{\rm ZTF}$ & $F_{{\rm var},r}$ (\%)\tablenotemark{a} \\
(1) & (2) & (3) & (4) & (5) & (6) & (7) & (8) & (9) & (10) \\
\midrule
J114515.94+794053.4 & High & 0.006 & 5.90 & 41 & 11 & 14.15 & 0.48 & 14.87 & 6.35 \\
J120309.60+443152.7 & High & 0.002 & 5.40 & 22 & 3 & 10.12 & 0.46 & 13.25 & 1.81 \\
J122548.87+333248.7 & High & 0.001 & 4.52 & 49 & 2 & 7.08 & 0.64 & 16.43 & 7.35 \\
J115628.17+550730.5 & Medium & 0.004 & 4.63 & 75 & 6 & 5.17 & 0.42 & 14.56 & 0.55 \\
J130141.56+100100.1\tablenotemark{b} & Medium & 0.027 & 5.34 & 23 & 2 & 2.85 & 0.54 & 17.14 & 0.99 \\
J125329.16+021006.0 & Medium & 0.003 & 5.56 & 23 & 3 & 2.55 & 0.27 & 13.92 & 0.36 \\
J083706.52+412722.8\tablenotemark{b} & Medium & 0.029 & 4.16 & 47 & 1 & 2.31 & 0.19 & 16.37 & 0.54 \\
J154610.43+125856.7 & Medium & 0.055 & 5.72 & 51 & 1 & 2.20 & 0.26 & 17.32 & 0.54 \\
J092046.28-080322.1 & Medium & 0.020 & 5.98 & 89 & 4 & 2.02 & 0.27 & 15.75 & 0.87 \\
J180143.84+445000.2\tablenotemark{c} & Medium & 0.041 & 5.85 & 81 & 9 & 2.02 & 0.28 & -- & 0.75 \\
J094908.76+022851.0 & Medium & 0.020 & 4.26 & 89 & 7 & 1.95 & 0.43 & 16.53 & 0.74 \\
J172005.27+535703.0 & Medium & 0.030 & 5.36 & 78 & 29 & 1.93 & 0.32 & 16.58 & 0.83 \\
J093249.50-050841.4\tablenotemark{b} & Medium & 0.023 & 4.74 & 89 & 5 & 1.85 & 0.33 & 16.71 & 0.24 \\
J110501.97+594103.6 & Medium & 0.034 & 6.80 & 14 & 6 & 1.84 & 0.36 & 15.60 & 3.50 \\
J132712.75+421754.4 & Medium & 0.042 & 5.19 & 49 & 4 & 1.82 & 0.37 & 16.07 & 0.61 \\
J103234.85+650227.9 & Medium & 0.006 & 5.52 & 48 & 6 & 1.81 & 0.38 & 15.61 & 0.82 \\
J132919.67+320017.1 & Medium & 0.024 & 5.85 & 50 & 5 & 1.78 & 0.35 & 15.97 & 0.85 \\
J085949.04+050339.1\tablenotemark{b} & Medium & 0.012 & 4.99 & 61 & 5 & 1.77 & 0.33 & 15.71 & 0.33 \\
J231154.80-001108.4 & Medium & 0.028 & 4.80 & 92 & 3 & 1.68 & 0.41 & 16.41 & 0.61 \\
J121031.22+402503.2 & Medium & 0.022 & 4.47 & 49 & 3 & 1.49 & 0.27 & 16.84 & 0.91 \\
\bottomrule
\end{tabular}
\par
\endgroup

\begingroup
\raggedright
\tablecomments{
(1): Source designation.
(2): Manual confidence of variability.
(3): Redshift.
(4): Black hole mass.
(5): TESS sector number with the largest $C_{\rm AMP}$ among those with $C_{\rm CORR}>0$ and without artifact flag.
(6): Number of available TESS sectors.
(7)--(8): Amplitude and correlation confidence for the
representative sector, as defined in
Section~\ref{ssec:metrics}.
(9): ZTF $i$-band PSF magnitude.
(10): Fractional rms variability amplitude of the ZTF
$r$-band light curve.
}

\tablenotetext{a}{
The long-term variability analysis is described in
Section~\ref{ssec:longterm}.
The photometry includes the reference-flux contribution, and
$F_{{\rm var},r}$ is not corrected for host-galaxy dilution.
}

\tablenotetext{b}{
These sources have neighboring stars or galaxies of comparable brightness within $40^{\prime\prime}$.
No Gaia DR3 variability classification is available for the examined neighbors, and our ZTF difference-image forced photometry does not establish intrinsic variability in them. These results do not rule out contamination of the TESS light curves.
}

\tablenotetext{c}{
No cataloged ZTF $i$-band PSF photometry is available;
the source satisfies host brightness screening with
$r_{\rm ZTF}=17.02$.
}
\par
\endgroup
\end{table*}

\subsection{Timescale analysis}\label{ssec:timescale}

\begin{figure*}[htbp]
\centering
\includegraphics[width=0.9\textwidth]{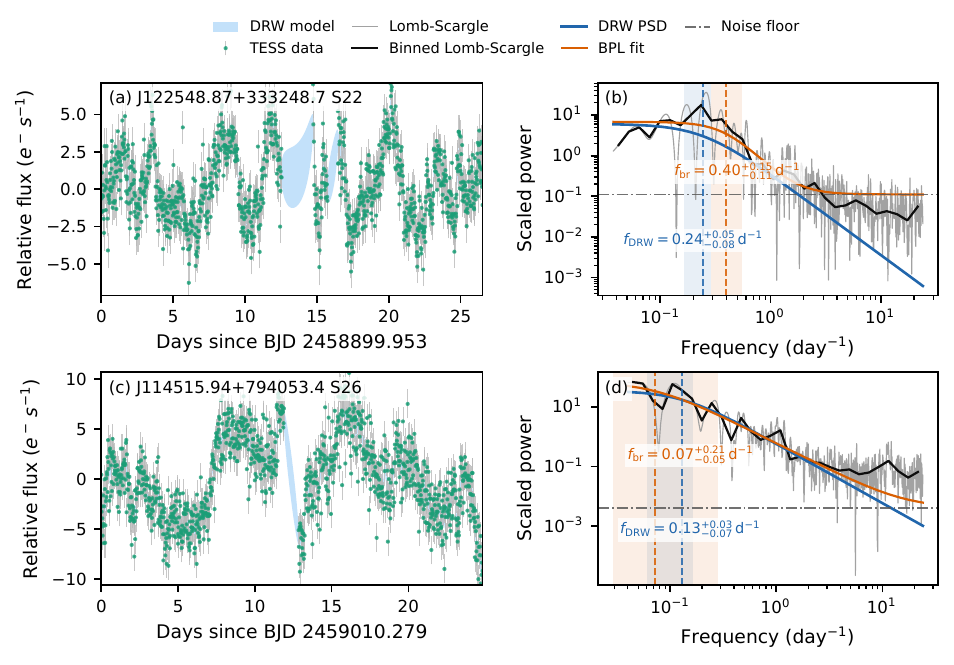}
\caption{
Example timescale analysis for high-confidence sectors.
Left panels show TESS light curves with flux errors. The blue shaded region shows the $1\sigma$ DRW predictive range.
Right panels show the Lomb--Scargle periodogram, its logarithmically binned form, the noise-free DRW PSD, and the empirical BPL fit.
Vertical dashed lines mark the DRW frequency ($f_{\rm DRW}$) and the BPL break frequency ($f_{\rm br}$), and the shading shows the $1\sigma$ uncertainty range.
\label{fig:timescale}}
\end{figure*}

The timescale analysis requires high-quality sectors with confident intrinsic variability and minimal instrumental artifacts. Figure \ref{fig:timescale} shows the source-aperture light curves and analysis results of two representative sectors. The DRW frequency uncertainties are estimated from posterior samples of the model fitting, while the BPL frequency uncertainties are estimated from the covariance of the weighted fit. The DRW frequency and the BPL break frequency are marginally consistent for J122548.87+333248.7 Sector 22. For J114515.94+794053.4 Sector 26, the BPL break frequency is also consistent with DRW, but it has large uncertainties. We note that the BPL fit is less robust for noisy sectors or for timescales close to the sector duration, as it has more free parameters. We therefore adopt the DRW result as the primary timescale estimate.

Compared to \cite{Burke20}, the damping timescale obtained for J122548.87+333248.7 (NGC 4395) Sector 22 is significantly shorter: for DRW, we obtained $\tau_{\rm DRW}=0.66^{+0.31}_{-0.11}\ {\rm days}$ while \cite{Burke20} reported $\tau_{\rm DRW}=2.3^{+1.8}_{-0.7}\ {\rm days}$; for BPL, we obtained $\tau_{\rm br}=0.40^{+0.15}_{-0.12}\ {\rm days}$ compared with the reported $\tau_{\rm br}=1.4^{+1.9}_{-0.5}\ {\rm days}$. This discrepancy likely reflects differences in light curve extraction and detrending. The full-hybrid \texttt{Quaver} reduction is designed to suppress trends common across the pixel grid, which may reduce sensitivity to the low frequency variability (see \citealt{Tripathi26} for a detailed comparison of the reduction modes).

\begin{figure}[h]
\centering
\includegraphics[width=0.45\textwidth]{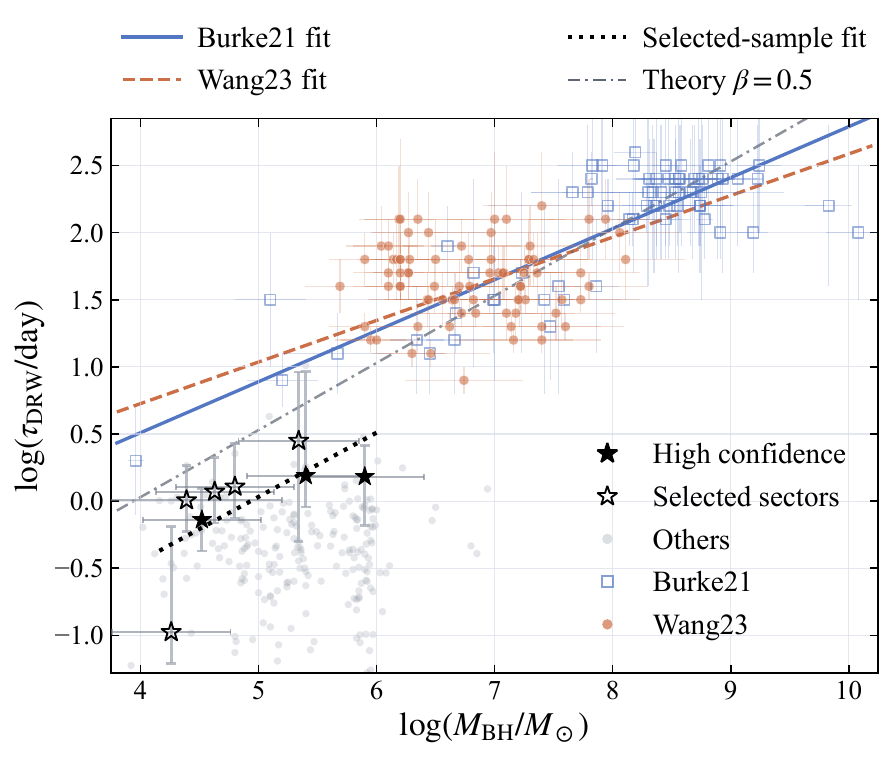}
\caption{
Correlation analysis between black-hole mass and DRW timescale. Literature measurements from \cite{Burke21, Wang23} are shown. The high confidence and the metric-selected sectors are included in this work's fitted correlation. A theoretical relation of $\tau_{\rm DRW}\propto M_{\rm BH}^{0.5}$ is shown alongside the fitted relations.
\label{fig:correlation}}
\end{figure}

For survey-level analysis, we tentatively select a subset of sectors with $C_{\mathrm{AMP}}>1.5$, $C_{\mathrm{CORR}}>0.4$, and a minimum usable coverage of 15 days. The threshold of $C_{\mathrm{AMP}}>1.5$ is approximated by the lowest amplitude that allows manual confirmation of a medium-confidence source; the $C_{\mathrm{CORR}}>0.4$ is to keep intrinsic variability as the primary driver of the light curve\footnote{The threshold is above many medium-confidence sectors because manual inspection may still distinguish intrinsic variability when instrumental artifacts are present, but DRW analysis generally cannot deal with these sectors.}. This selection retains 17 sectors from 8 unique sources. One of these sources, J101848.59+382810.7, has low manual confidence and is retained by the quantitative selection. As shown in Figure \ref{fig:correlation}, we take the source-median $\tau_{\rm DRW}$ of available sectors and fit the correlation between $\tau_{\rm DRW}$ and $M_{\rm BH}$, and we compare it with the more established correlation for higher mass AGNs \citep{Burke21, Wang23}. The timescales of this work are reported in the observer frame; the rest-frame correction is \(<0.03\) dex for this low-redshift sample and does not affect the qualitative result. Note that the mass estimates adopted from \cite{Guo26} and \cite{Pucha26} may differ from the value reported in \cite{Burke21} and \cite{Wang23}.

Our DRW estimates are generally lower than the best-fit lines obtained in previous studies by 0.5--1.0 dex, although established timescale measurements below $10^6M_{\odot}$ are sparse. The 8 sources yield Spearman's rank correlation $\rho=0.88$ ($p=0.007$) and Pearson's correlation $r=0.65$ ($p=0.066$) in log space, using two-sided exact permutation tests, with an unweighted linear slope of $\beta=0.48$. This slope is steeper than the $\beta=0.38$ relation found by \citet{Burke21} and the $\beta=0.31$ relation adopted by \citet{Wang23}, and is closer to the standard accretion-disk expectation of $\tau_{\rm DRW}\propto M_{\rm BH}^{0.5}$ discussed by \citet{Burke21}.

However, we must note this tentative correlation is very sensitive to the sample threshold and cannot be established from the full TESS dataset. Varying the $C_{\mathrm{AMP}}$ threshold from 1.0 to 2.0 yields slopes of $\beta=0.24$--$0.58$, while varying the $C_{\mathrm{CORR}}$ threshold from 0.3 to 0.4 yields slopes of $\beta=-0.13$ to $0.48$.
The gray points in Figure \ref{fig:correlation} show the DRW results of sectors that do not satisfy the threshold ($C_{\mathrm{AMP}}>1.5$ and $C_{\mathrm{CORR}}>0.4$), which generally correspond to weaker intrinsic variability. This broad sample shows no evidence of the correlation (Spearman $\rho=-0.005$). This suggests that, as the variability metrics decrease, noise and residual artifacts become increasingly dominant in DRW measurements, smearing out the physical signal of the \(\tau_{\rm DRW}\)--\(M_{\rm BH}\) relation. Regardless of sector quality, almost all DRW timescales obtained in this work fall below the literature best-fit lines. This suggests that, in DRW analyses, TESS light curves reduced with Quaver in full-hybrid mode may systematically differ from those used in most studies based on lower-cadence observations. 

\subsection{Long-term variability}\label{ssec:longterm}

\begin{figure*}[htbp]
\centering
\includegraphics[width=0.9\textwidth]{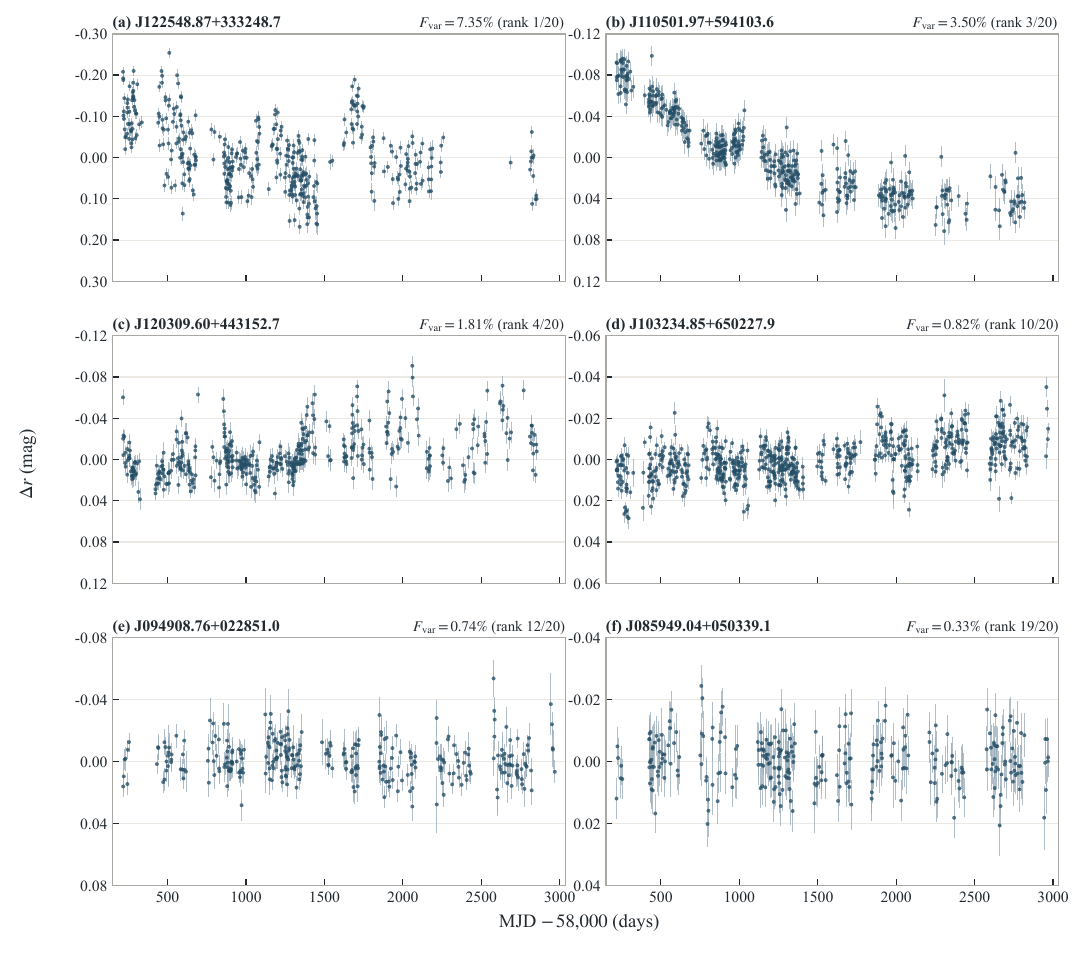}
\caption{Representative ZTF $r$-band long-term light curves spanning the $F_{\rm var}$ distribution of the high- and medium-confidence sources. Panel labels give $F_{\rm var}$ and its rank within the 20-source sample.
\label{fig:longterm}}
\end{figure*}

Assuming the variability-selected sources have intrinsic short-term variability on TESS baselines, their long-term variability could be detected as well. We followed the approach in \cite{Arevalo26} to examine the 20 high- and medium-confidence sources with ZTF data, performing forced photometry on difference images to obtain AGN light curves at lower noise level.

The fractional variability amplitudes ($F_{\mathrm{var}}$) are reported in Table \ref{tab:high_medium_candidates}. The high-confidence sources show long-term $F_{\mathrm{var}}$ ranging from 1.8 to 7.4\%; most of the medium-confidence sources have $F_{\mathrm{var}}$ below 1\%, with the exception of J110501.97+594103.6 having $F_{\mathrm{var}}=3.5\%$ and a long-term fading trend. This source has discrepant black hole mass estimates from \citet{Greene07} ($\log(M_{\rm BH}/M_\odot)=5.5$) and \citet{Guo26} ($\log(M_{\rm BH}/M_\odot)=6.8$), possibly due to different broad--narrow line decomposition in spectral fitting; it has been tentatively identified as a tidal disruption event (TDE) candidate \citep{Wang22}. Figure \ref{fig:longterm} shows ZTF r-band light curves of representative sources spanning the range of $F_{\mathrm{var}}$ in the sample. Statistics based solely on formal photometric errors may overstate the variability detection confidence \citep{Masci23}. We therefore do not regard the low-amplitude variations of most medium-confidence sources as reliable evidence of long-term variability.

\section{Discussion and Conclusions}\label{sec:conclusion}

We have presented a TESS variability search for IMBH AGN candidates using \texttt{Quaver}-reduced light curves. We develop variability metrics based on source-to-background amplitude contrast and light curve coherence, together with diagnostics designed to screen for instrumental artifacts and neighboring source contamination. This procedure identifies 3 high-confidence and 17 medium-confidence variable candidates. The high-confidence sources are known nearby AGNs with previously reported optical variability, while most medium-confidence candidates have not been well characterized and are therefore useful targets for follow-up observations.

For the TESS sectors with high intrinsic variability, we modeled the light curves with a damped random walk (DRW) and found a very tentative positive relation between the damping timescale and black-hole mass with $\beta=0.48$. This steeper slope is consistent with baseline-corrected DRW results \citep{Xia25} and recent frequency domain studies \citep{Arevalo24, Petrecca26}, and \cite{Su24} also suggested a relatively steep slope in the IMBH mass range. However, the limited number of fitted sectors, photometric sensitivity, and observing baselines leave this slope poorly constrained. With a limited baseline of 28 days, long DRW timescales might be biased downward \citep{Kozlowski17}. We performed injection--recovery tests with the same baseline and cadence as real observations, which show that DRW timescales beyond 3 days are increasingly underestimated (Appendix~\ref{App:recovery}). This could contribute to systematic timescale underestimation of some sources with $M_{\rm BH}$ close to $10^6\,M_\odot$. In addition, the DRW-based variability timescale analysis should be interpreted cautiously because the DRW model itself may not perfectly describe AGN variability, especially at very low frequencies \citep{Guo17} or at the high frequency end \citep{Su24}. Therefore, the $\tau_{\rm DRW}$ in this work should be cautiously interpreted as the timescale sampled by TESS baseline rather than a unique physical timescale.

The TESS measurements probe variability on hour-to-month timescales with its high cadence, broad sky coverage, and repeated observations over multiple missions, which is potentially useful for identifying IMBH candidates with intermittent or short-timescale variability. At the same time, TESS is limited by the large pixel scale and PRF which risk contamination from neighboring sources, the relatively shallow limiting magnitude, and the sector-level artifacts that can cause fake variability signals.

Future survey programs can detect and model variability timescales with promising capabilities. Rubin/LSST \citep{Ivezic19} will offer much deeper fields over a decade-long baseline, making it valuable for confirming long-term variability for faint IMBH AGNs, although its temporal sampling is much sparser than that of TESS. Closer space-based analogs to TESS are PLATO \citep{Rauer25} and Earth 2.0 (ET; \citealt{Ge24}), which will provide high-cadence photometry over selected large fields for baselines much longer than a typical TESS sector. For AGNs that fall within its long-duration fields, PLATO could extend the optical variability baseline toward hour-to-year timescales, providing a stronger test of DRW timescales in the IMBH regime.

\begin{acknowledgments}

WWZ is supported by the Shanghai Natural Science Foundation Youth Project (Grant No. 25ZR1402546), Strategic Priority Research Program of the Chinese Academy of Sciences (Grant No. XDB0800302) and the National Key Research and Development Program of China (Grant No. 2025YFA1614102).
A.T. acknowledges support from the Xinjiang Tianchi Talent Program,   the Tianshan Talent Training Program (Grant No. 2023TSYCCX0099) and the National 
Key R\&D Program of China (Grant Nos. 2024YFA1611503 and 2022SKA0120102).

\end{acknowledgments}

%
\facilities{TESS, ZTF, Gaia}

\software{\texttt{Quaver} \citep{Quaver23}, \texttt{Astropy} \citep{Astropy22}, \texttt{Lightkurve} \citep{Lightkurve18}, \texttt{EzTao} \citep{EzTao}, \texttt{emcee} \citep{Foreman13}, \texttt{Matplotlib} \citep{Matplotlib07}, \texttt{SExtractor} \citep{Bertin96}
          }


\appendix

\section{TESS artifact quality flag}\label{App:flag}

\begin{figure*}[htbp]
\centering
\includegraphics[width=0.8\textwidth]{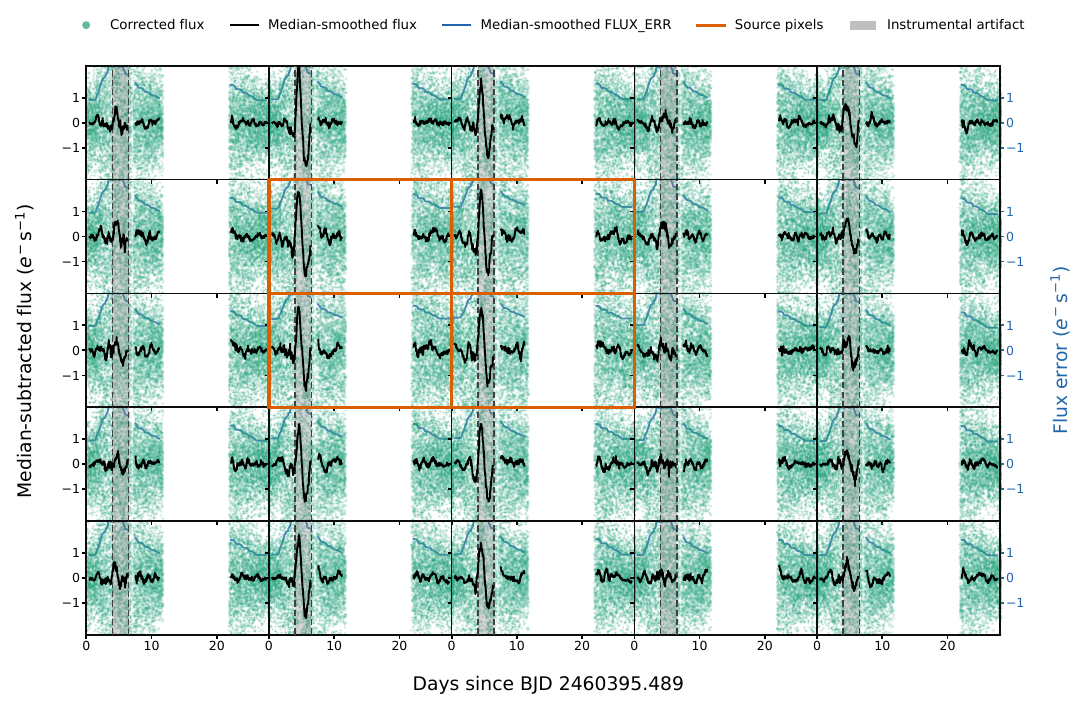}
\caption{Example \(5\times5\) pixel-grid product for J162549.67+480601.2 in Sector 77 after \texttt{Quaver} reduction. Green points show corrected fluxes, whereas black and blue curves show the 0.5-day median-smoothed flux and flux error. The four source pixels are outlined in orange, and the spike-shaped instrumental artifact is shaded in gray.
\label{fig:Flux_spike}}
\end{figure*}

We flag the TESS sectors potentially affected by instrumental artifacts by utilizing flux error information, since the artificial variability often occurs at periods of elevated flux error (see Figure \ref{fig:Flux_spike} as an example). Specifically, we examine the following two aspects: 1) does the flux error vary significantly within a sector, and 2) does the apparent variability occur at periods of elevated flux error. 

We quantify the first aspect by computing the 0.5-day median flux error across the $5\times5$ pixel grid and take the ratio of the 95th and 10th percentiles of flux error, denoted as $E_{\mathrm{err}}$. For the second aspect, we compute the absolute deviation from median and define the flux-error overlap fraction, which measures how often the top 10\% deviations coincide with the top 10\% error cadences, denoted as $F_{\mathrm{overlap}}$. For each of the two computed metrics, we selected cutoffs near the top 20\% percentile ($E_{\mathrm{err}}=1.7$, $F_{\mathrm{overlap}}=0.35$). Then, we excluded the sectors that satisfy both criteria, flagging a total of 101 (6.9\%) sectors.

\section{DRW timescale injection--recovery tests}\label{App:recovery}

To assess the effects of limited observing baselines and residual noise on timescale analysis, we performed injection--recovery tests of DRW timescales. We followed the baseline, cadence, and DRW-fitted amplitude of the 17 sectors from 8 sources selected in Section~\ref{ssec:timescale}. For each sector,
we generated 100 stationary DRW realizations at each input timescale $\tau_{\rm input}=0.1$, 0.3, 1, 2, 3, 5, 10, and 20 days. For injected noise, we tested both Gaussian noise with amplitude equivalent to nominal flux error and observed residual noise obtained by summing four-pixel apertures in background pixels.

\begin{figure*}[htbp]
\centering
\includegraphics[width=0.45\textwidth]{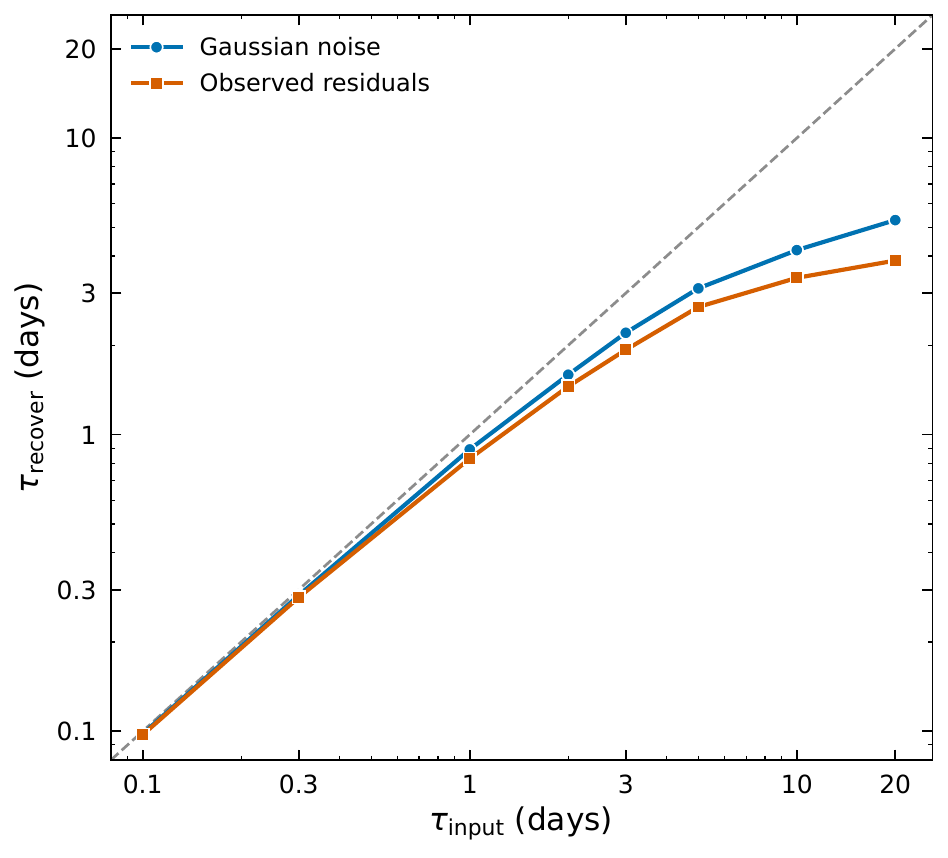}
\caption{Median input and recovered DRW timescale for Gaussian noise and observed residual noise. The dashed line indicates perfect recovery.
\label{fig:recovery}}
\end{figure*}

Figure~\ref{fig:recovery} shows the median recovered timescales for both types of injected noise. At $\tau_{\rm input}=3$ days, the median recovered timescales were 2.20 days with Gaussian noise and 1.93 days with observed residuals; at $\tau_{\rm input}=20$ days, they were 5.29 and 3.86 days, respectively. These tests suggest that long intrinsic DRW timescales could be significantly underestimated in the DRW analysis.

\bibliography{sample701}{}
\bibliographystyle{aasjournalv7}



\end{document}